\documentclass[showkeys,showpacs,amsmath,amssymb,groupaddress,twocolumn,nofootinbib,prl]{revtex4-1}

\usepackage{graphicx}
\usepackage{subfig}
\usepackage{amssymb}
\usepackage{color}
\usepackage{mathrsfs}
\usepackage{float}
\usepackage{mathtools} % needed for declaring paired delimeters (bras and kets)
\usepackage{bbold}

\DeclarePairedDelimiter{\bra}{\langle}{\rvert} %needs package mathtools
\DeclarePairedDelimiter{\ket}{\lvert}{\rangle}

\newcommand{\ketbra}[1]{\ket{#1}\bra{#1}}
\newcommand{\ketbraa}[2]{\ket{#1}\bra{#2}}

\newcommand{\ddd}{\textrm{d}}

\newcommand{\ii}{\textrm{i}}
\newcommand{\tp}{t^{\prime}}

\DeclareMathAlphabet{\mathpzc}{OT1}{pzc}{m}{it}

\begin{document}

\newpage

\title{Numerical Feynman integrals for density operator dynamics using master equation interpolants: faster convergence and significant reduction of computational cost}
%\title{Numerical Feynman integrals with physically inspired interpolation: faster convergence and significant reduction of computational cost}
\author{Nikesh S. Dattani}

\vspace{100mm}

\affiliation{Hertford College, University of Oxford, Oxford UK, OX1 3BW}

\date{\today}

\begin{abstract}
%The Feynman integral is one of the most accurate methods for calculating density operator dynamics in open quantum systems. However, the number of time steps that can realistically be used is always limited, therefore one often obtains an approximation of the density operator at a sparse grid of points in time. Instead of relying only on \textit{ad hoc} interpolation methods such as splines to estimate the system density operator in between these points, I propose a method that uses physical information to assist with this interpolation. This method is tested on a physically significant system, on which its use allows important qualitative features of the density operator dynamics to be captured with as little as 2 time steps in the Feynman integral. This method allows for an enormous reduction in the amount of memory and CPU time required for approximating density operator dynamics within a desired accuracy. Since this method does not change the way the Feynman integral itself is calculated, the value of the density operator approximation at the points in time used to discretize the Feynamn integral will be the same whether or not this method is used, but its approximation in between these points in time is considerably improved by this method.
The Feynman integral is one of the most accurate methods for calculating density operator dynamics in open quantum systems, but due to its enormous computational cost, one can only use it to calculate an approximation of the density operator at a sparse grid of points in time. Conventionally, interpolation methods such as splines are then used to to estimate the density operator in between these points, but this can lead to serious problems such as the loss of positivity. In this work a method is presented which uses physical information about the system to improve this interpolation. The method is tested on a physically significant system and allows for a huge reduction in the amount of memory and CPU time required. 
\end{abstract}


%03.65.Yz=Decoherence; open systems; quantum statistical methods (see also 03.67.Pp in quantum information; for decoherence in Bose-Einstein condensates, see 03.75.Gg)
%02.70.-c=Computational techniques; simulations (for quantum computation, see 03.67.Lx; for computational techniques extensively used in subdivisions of physics, see the appropriate section; for example, see 47.11.-j Computational methods in fluid dynamics)
%02.60.Ed=Interpolation; curve fitting

\maketitle

\section{Introduction}
%[One of the most important properties of a quantum mechanical system is its density operator (matrix).  (well Wigner distribution's just as good)]

%The most informative way to describe a quantum mechanical system is by its density operator $\rho$. 
The statistical properties of a quantum state can be uniquely described by its density operator $\rho$.
We are often interested in how the density operator of an \textit{open} quantum system (OQS) evolves in time, and an enormous amount of effort has been dedicated towards developing mathematical techniques for predicting this evolution in time.

To date, one of the most powerful of these techniques is based on the Feynman integral\cite{1948Feynman} and the Feynman-Vernon influence functional\cite{1961Wells,*1963Feynman,*1965Feynman}. This method, whose computation essentially only relies on integration, has substantial advantages over methods which rely on differential equations for the density operator (quantum master equations). 

%This is because numerical integrations can be done very systematically, while there's no orderly way of approaching a general DE (quantum master equations are often partial, non-local, sometimes stochastic, integro-differential equations that can't be solved without approximations that corrupt the solution.

%wait is that the real reason ? or is it because what yuti said about the infinite bath ? I think differential equations can be 'approached' a little bit systematically (althought not easy), the primary problem is DERIVING the master equation int he first place ? otherwise you get what yuti said about the infinite dimensional environmental hamiltonian, and in the path integral you just integrate out the bath.

Unfortunately, although the Feynman integral methods are conceptually straightforward, the cost of their numerical evaluation often increases exponentially in the number of time steps used. Contrarily, very small time steps are required in situations where $\rho$ varies quickly.

 %To reduce the computation time, one can store the value of the influence functional and the unitary evolution functional (objects whose values are reused in some form at each iteration) over each relevant path. 

In fact, with current technology, using these methods is often intractable for the results desired.
To reduce the computation time, one can store numbers that are reused in more than one iteration. But merely storing the amplitude over each Feynman path ordinarily requires an array of $M^{2(N+1)}$ complex-valued elements ($M$ = dimension of the hilbert space of the OQS, or of the number of points in the discretization of the system coordinate in the case of an infinite dimensional hilbert space, and $N$ = number of time steps, excluding $t=0$, elapsed before the influence functional becomes constant with respect to time). For a two-level system (2LS) with $N=20$, this amounts to $>70$TB\footnote{Throughout this paper, the SI convention will be used, so 1TB=10$^{12}$ bytes rather than the binary convention in which 1TB=1TiB=2$^{40}$ bytes.} of memory (assuming double precision arithmetic). If $N$ is increased to 25, this becomes $>72$PB\footnote{At the top of the current TOP500 list of supercomputers is the Cray XT5 Jaguar which has hard drive space on the order of 10PB\cite{2009Bland}}. 

The amount of memory required can be significantly reduced without too big of an impact on the accuracy, by filtering out some of the smaller Feynman amplitudes\cite{1996Sim,*2001Sim}, but despite such filtering, there is always a maximum $N$ (henceforth denoted $N_{\textrm{max}}$)beyond which there is no longer enough memory for the surviving amplitudes to be stored, and the remaining ones would have to be recaluclated at each iteration (an excruciatingly time-demanding process). 

The point at which $N$ can no longer practically be increased, and still does not meet the demands of the situation, occurs very often in real systems of interest. Even in the simple case of a  driven 2LS, when the driving frequency becomes large enough, using the maximum computationally feasible $N$ will lead to an approximation of $\rho$ at a grid of points which is too sparse for typical interpolation techniques to capture important features such as peaks and troughs. 

%In these situations, apart from abandoning the Feynman integral approach, the only option has been to settle for using the maximum computationally feasible $N$. This would lead to an approximation of the density matrix at a finite number of points in time
 
This paper explores a method that attempts to capture these features, without increasing the amount of memory required. Instead of interpolating between these sparse points using conventional techniques, the interpolation incorporates some of the information about the relevant physics of the system.

\section{Example calculations}

All example calculations below will be for a 2LS with hamiltonian: $H=\frac{1}{2}\frac{5\pi}{4}(\ketbraa{0}{1}+\ketbraa{1}{0})$, and spectral density: $J(\omega)=0.0268~\omega^3e^{(-0.2063\omega^2)}$ at a temperature of 25K. This hamiltonian corresponds to a GaAs quantum dot being driven by a pulse with frequency $\Omega=\frac{5\pi}{4}$, and the spectral density is from a very recent experimental study\cite{2010Ramsay2}. The density matrix in every case will be initialized at $\rho=\ketbra{0}$, and its evolution will be calculated using the QUAPI as described in \cite{1995Makri,*1995Makri2} for a duration of 3.5ps. For this entire duration, the influence functional for this spectral density never settles with respect to time, so the markovian iteration method described in \cite{1994Makarov} will not be used.

I will also assume that the amplitude over each Feynman path is stored in an array named $A$, rather than being recalculated at each time step. For pedagogical reasons, no filtering of these amplitudes is performed in this paper, but the methods described here can easily be combined with a filtering process such as \cite{1996Sim} or \cite{2001Sim} to save computational cost.

\section{The method}

%\subsection{Pedagogical assumptions and example system}

Conventionally, the numerical calculation of a Feynman integral involves dividing the time axis into a discrete mesh. As explained above, the number of points on this mesh is $\le$ some $N_{\textrm{max}}$ (this excludes the starting point $t=0$). Let's define this mesh the \textit{Feynman mesh}.

 Two calculations are shown below for the example system described above, each with a different number of time steps $N$. The square-shaped markers indicate the points on the Feynman mesh for each specific curve, and the points are joined by cubic splines (by the SPLINE function in MATLAB2010a).
 
\hspace{-5mm}\includegraphics[width=0.48\textwidth]{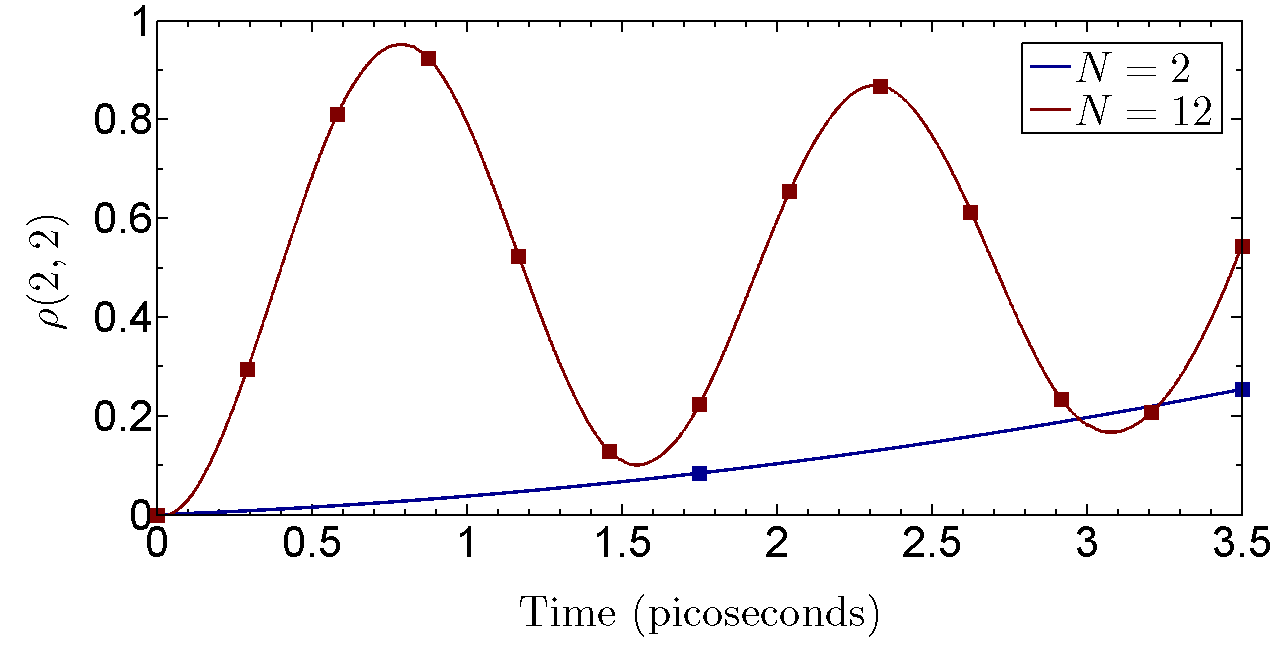}

It's clear that with $N=2$, some important qualitative features are missing. Unfortunately, some physically interesting quantum mechanical systems have a large enough hilbert space that using $N$ as low as 3 would currently be unachievable on a common computer. One such example is the hilbert space of single excitations in the the light-harvesting system LH2, the dimension of which is 27\cite{1997Fleming}.

For such cases, no conventional interpolation technique would be able to capture qualitative features such as in the figure above. We need a better way to approximate $\rho$ at places between the points of the Feynman mesh.

To do this, let's construct a finer mesh, for example:

\includegraphics[width=0.45\textwidth]{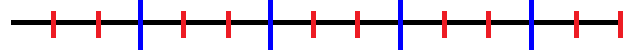}.

Here, the blue (coarser) mesh is the Feynman mesh - for each point on this mesh, the size of $A$ increases by a factor of $M^2$. The ambition is to approximate $\rho$ at the points on the red (finer) mesh, in a manner that not only resembles the true physics, but also does not increase the number of amplitudes over which to sum.

There is a very simple way to do this. On the points of the Feynman mesh, $\rho$ is still calculated exactly as before, so the value of $\rho$ at these points is in no way damaged and this calculation can be completed before anything else. Then we can use a master equation to approximate $\rho$ in between these points. A simple example is the von Neumann equation for a closed quantum system with the same system hamiltonian. If the hamiltonian is time-independent, this is:

\begin{equation}
\rho(t+n\Delta t)=U^n \rho(t) {U^{\dag}}^n ~~~ , ~~~ U=e^{-\ii H\Delta t},
\label{eq:vonNeumannIterations}
\end{equation} 
 
\noindent
where $\Delta t$ is the size of the time-steps on the finer of the two meshes, and $t$ is a point in time that is on the Feynman mesh. For time-dependent hamiltonians, this becomes:

\begin{equation}
\rho(t+n\Delta t)=e^{-\ii\int_t^{n\Delta t} H(\tp) \ddd \tp}\rho(t) e^{\ii\int_t^{n\Delta t} H(\tp) \ddd \tp}.
\label{eq:vonNeumannIterationsTD}
\end{equation} 

%Henceforth, calculations will be labeled according to the structure of the time discretization. A curve labeled QUAPI-DM$(N,NN)$ QUAPI-DM-

%In the time discretization in the figure above, the number of points on the Feynman mesh is $N=5$, and the number of points in each Feynman interval is $NN=2$. We can first calculate $\rho$ at each of the 5 points on the Feynman mesh, 
%
%we approximate $\rho$ by 

The figure below shows that when $\rho$ is approximated at the points of the finer mesh in this way, the qualitative features of a QUAPI calculation with $N=12$ can be recovered quite impressively by a calculation with only $N=2$, provided that the number ($NN$) of points on the finer mesh lying between points of the Feynman mesh is increased to 2. The meshes for the upper diagram were designed such that the three points on the $N2NN0$ Feynman mesh also happen to be points on the Feynman mesh of all other cases (the total number of points on the mesh is $1+(NN+1)N$). The meshes for the lower diagram use the maximum number of evenly spaced points that can fit in the given time interval ($1+(NN+1)(N+1)$), so the size of the time steps for a given pair $(N,NN)$ is reduced in order to fit an extra $NN$ points. The solid black curve was calculated with $N12NN0$, and is the accurate benchmark to which all other calculations in this paper are compared. The dotted black curve is what the evolution would look like if the OQS was closed (ie, unitary dynamics according to the von Neumann equation without any system-environment interaction).

\includegraphics[width=0.45\textwidth]{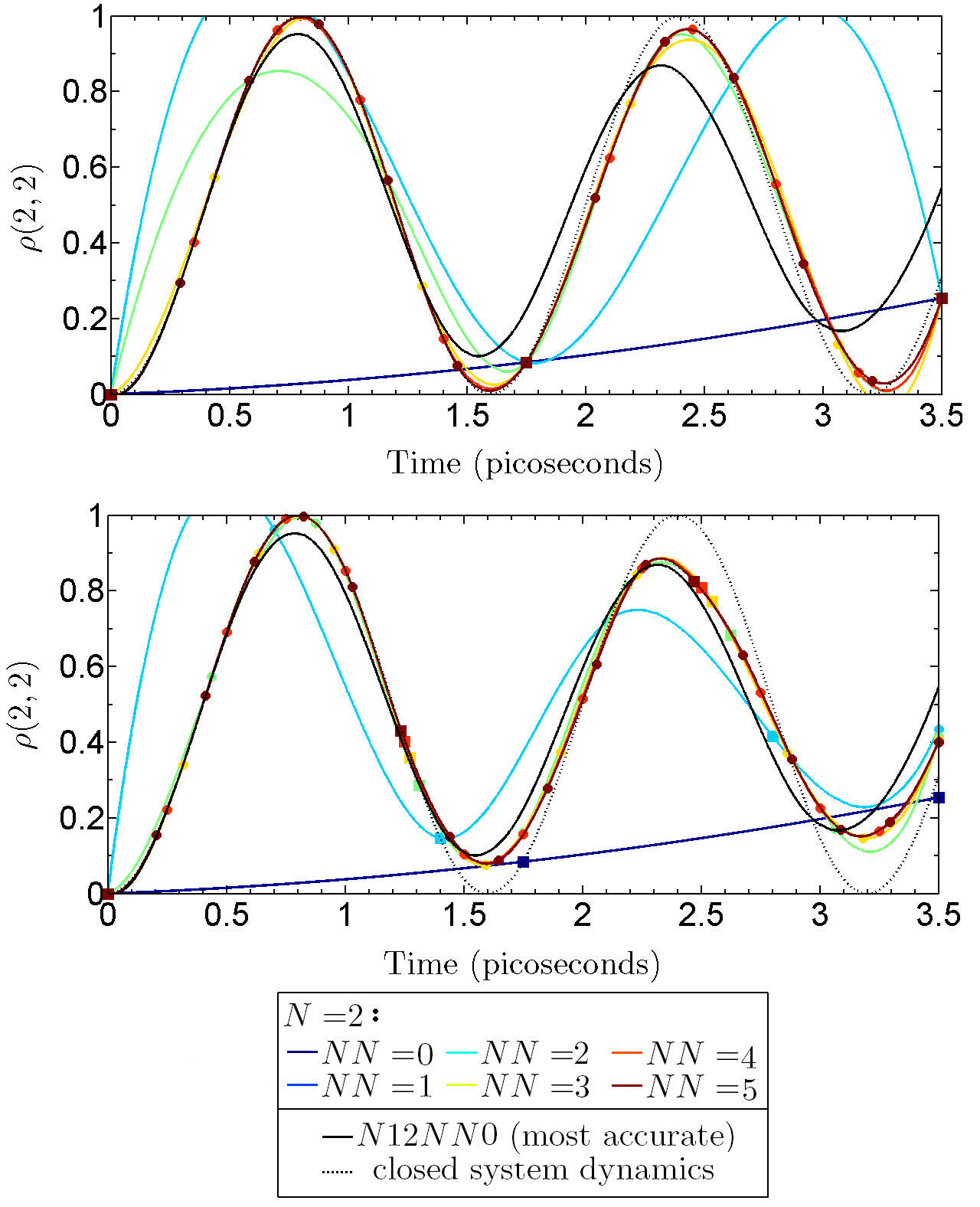}

In the upper diagram we can see that all curves with $N=2$ have two points in common, which are the points on the Feynman mesh. This illustrates the fact (that was stated earlier), that the calculated value of $\rho$ at the points of the Feynman mesh are in no way affected by this interpolation technique. The curves are simply clamped by the points on the Feynman mesh, and in between those points they evolve as if there was no interaction between the system and the environment. Because $\rho$ is not approximated very well at the points of the $N=2$ Feynman mesh, all of the curves are clamped at somewhat inaccurate values, and therefore, regardless of how accurate the unitary approximation is on the finer mesh, some features such as the maxima and minima are not captured very well.

In the lower diagram, the curves are clamped at different points because the size of the time steps on the Feynman mesh is reduced as $NN$ increases. As a result, the maxima and minima are approximated slightly better. In fact, if we allowed the points on the Feynman mesh to be distributed unevenly, the curves could be clamped where the minima are predicted to be, and with $N$ still equal to  2, we could get results that look even better than those  in the lower diagram. But to keep this study simple, all results in this paper will use meshes with evenly spaced points.

%It is important to note that 

\section{Faster convergence}

When a Feynman integral is numerically approximated, its accuracy is usually checked by systematically increasing the number of time steps and monitoring the results until they have converged. Once increasing the number of time steps leads to no change in the result (up to some tolerance), it is believed that the converged result is accurate (provided that the error in all approximations, such as the Trotter splitting of the time evolution operator, are only dependent on the size of the time step - this is usually true).

The diagram below shows that for larger values of $NN$, the curves converge quicker as $N$ is increased. The size of the double-precision array $A$ and the approximate CPU time for each calculation are shown in the legend - neither of these quantities changed at all as $NN$ was varied.

\includegraphics[width=0.45\textwidth]{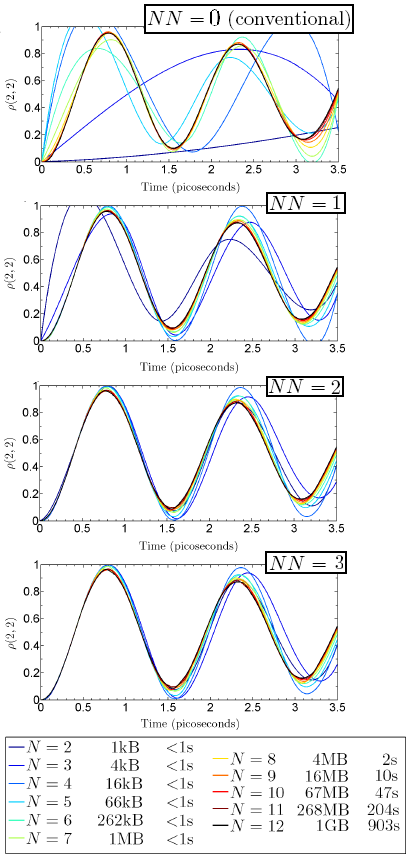}

\section{Accuracy}

We have looked at diagrams in which $N$ was fixed and $NN$ varied, and diagrams in which $NN$ was fixed and $N$ was varied. It is certainly apparent from fig. 2 that increasing $NN$ allows calculations with low $N$ to capture qualitative features that wouldn't be captured conventionally (with $NN=0$). But since $N$ is only 2, and therefore $\rho$ is not clamped very accurately at the points on the Feynman mesh, this diagram doesn't convince us that a calculation with a high $N$ can be accurately reproduced by using a lower $N$ with a higher $NN$.

It is also certainly apparent from fig. 3 that by increasing $NN$, the calculations converge up to a reasonable tolerance with a lower $N$, but since each diagram uses a different $NN$, it is not obvious whether or not the curves in every case are converging to the (most reliable) converged result of the $NN=0$ diagram.

The figure below shows the most reliable converged result ($N12NN0$), and compares it to results for $N10NN1$, $N8NN1$ and $N6NN1$. For the latter three curves, the dotted lines represent their corresponding $NN=0$ results; and the  $NN=2$ and 3 curves were not displayed because for these $N$ values, those curves look essentially the same as the $NN=1$ curves. 

\includegraphics[width=0.45\textwidth]{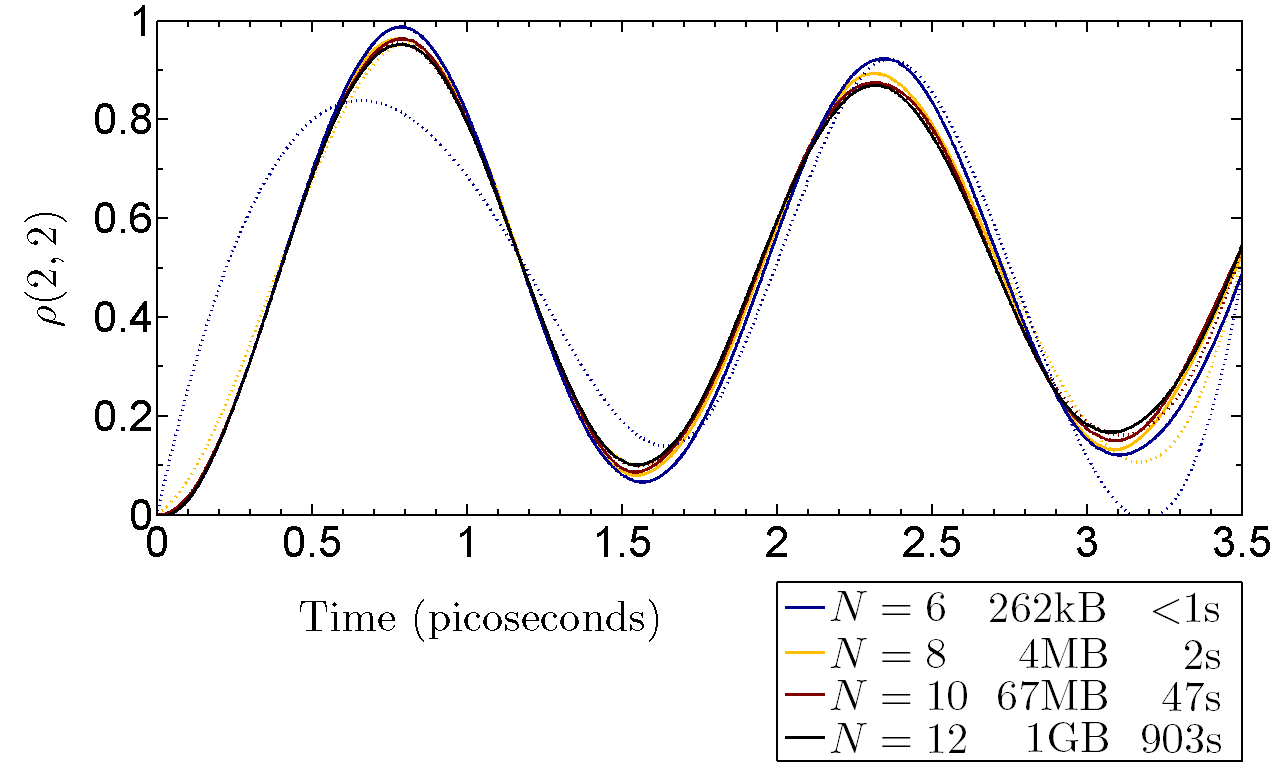}

%use N6,8,10,12,NN1&2.png not N6,7,10,12,NN1&2.png because there's a part in which the NN=1 case looks BETTER than the NN=2 for N=8

This figure shows that with $NN=1$, the $N12N0$ result can quite accurately be reproduced with only $N=6$, and even more accurately with $N=8$. More importantly, it shows that calculations done the conventional way ($NN=0$) for the same two cases differ very largely from the desired result. This outcome can be extremely useful in cases where the demands of the calculation can not be met by the computational resources available (a very common situation for numerical Feynman integrals). 

For example, if one desires to predict the decoherence rate of an entangled state of two qubits (a quantity that's tremendously useful when choosing material parameters for building a quantum computer), the dimension of the Hilbert space is 4, so just storing the array of double precision amplitudes for 10 time steps would require 281TB of memory if no filtering is done or if filtering doesn't help. This is pragmatically impossible today, but if we assume that the result in the above figure doesn't completely deteriorate when applied to the considered system, a remarkably accurate approximation can be calculated with only 6 time steps. The equivalent array would only require 4.3GB, and would therefore easily fit in the RAM of a fairly good modern laptop computer.

One more observation from fig. 4 is that although both $N=10$ curves lie closer to the converged result than the $N8NN1$ result, there is not much of a difference between the $NN=0$ and $NN=1$ cases, simply because the $NN=0$ case already seems to have enough points. It is important to realize that increasing $NN$ will not always improve the accuracy of a calculation, especially when the size of the time intervals in the Feynman mesh are already very small in comparison to the time scale on which $\rho$ changes.

\section{Discussion}

One feature of the method described above that is subideal is that during the unitary evolution of $\rho$ which occurs between the points of the Feynman mesh, no information about the influence of the environment on the OQS is used. This could very easily be improved - instead of using the solution of the von Neumann equation for a closed system to propagate $\rho$ in these time intervals, one could use a more sophisticated master equation that incorporates information about the influence of the environment.

Another feature that is independently subideal, is that even though $\rho$ is being approximated at points in between the points of the Feynman mesh, no information is being used to try to update the amplitudes during these time intervals. Consequently, every approximation of $\rho$ on a point of the Feynman mesh is only as accurate as it would have been (with a Feynman mesh of the same size) if this method was not used at all. 

For example, the interpolation technique described above neither improves nor impairs the accuracy of plots where the only points considered are on the Feynman mesh, such as ones where the elements of $\rho(t=15\textrm{ps})$ are plotted as a function of driving frequency. If $t=15$ps happens to be on the Feynman mesh, absolutely no improvement is
 made in the approximation of $\rho(t=15$ps$)$. 
%such as ones where the elements of $\rho(t)$ after a chosen amount of time are plotted against the driving frequency. If the chosen duration happens to be a point on the Feynman mesh, absolutely no improvement is made to the approximation of $\rho$
Even if the size of the Feynman time intervals were shortened without changing $N$ (in the same way they are shortened from the fig. 2a to fig. 2b), the improvement this method would provide would be negligible if $t$ is large enough, which would usually be the case.

There is more than one way in which physical information can be used to update the amplitudes more often without increasing the size of the array containing them. For one, we could continue to update $\rho$ in a manner resembling eq. \eqref{eq:vonNeumannIterations}, but simultaneously use the operators in eq. \eqref{eq:vonNeumannIterations} to update the elements of $A$. Alternatively, instead of using \eqref{eq:vonNeumannIterations} to update $\rho$ on the points of the finer mesh, we could update these points by calculating a Feynman integral, except instead of increasing the number of amplitudes by a factor of $M^2$ (as would be done at each point on the Feynman mesh), the summation could be done on updated versions of the amplitudes left over from the previous point on the Feynman mesh. These updated amplitudes would ideally contain information from both the closed system hamiltonian and from the influence functional. One could either use the influence functional from the previous Feynman point to do this, or one could actually update the influence functional along the way.

These enhancements to the method described in this paper will be investigated much more thoroughly in future work.

\bibliographystyle{abbrv}
%\bibliography{C:/Users/Nikesh/Desktop/Research/bib}

\end{document}